\documentclass[11pt,a4paper,twocolumn]{article}
\usepackage[top=1.0cm, bottom=1.5cm, left=1.5cm, right=1.5cm, a4paper]{geometry}
\usepackage{authblk}
\usepackage[utf8]{inputenc}
\usepackage[english]{babel}
\usepackage[leftcaption]{sidecap}
\usepackage{url}
\usepackage{graphicx}
\usepackage{cite}

\usepackage{xcolor}
\newcommand{\yl}[1]{\textcolor{black}{#1}}

\makeatletter
\renewcommand\AB@affilsepx{, \protect\Affilfont}
\makeatother

\let\OLDthebibliography\thebibliography
\renewcommand\thebibliography[1]{
  \OLDthebibliography{#1}
  \setlength{\parskip}{0pt}
  \setlength{\itemsep}{0pt plus 0.3ex}
}

\title{
Software Defined Radio for on-line interaction with beam processes in the heavy ion storage ring ESR}

\author[1]{M.~S.~Sanjari}
\author[1, 2]{Yu.~A.~Litvinov}
\author[1]{S.~Litvinov}
\author[1]{B.~Peter}
\author[3]{R.~J.~Chen}
\author[4]{D.~Dmytriiev}
\author[1]{C.~Forconi}
\author[1]{J.~Glorius}
\author[5, 6]{G.~W.~Hudson-Chang}
\author[1]{H.~Hüther}
\author[7]{E.~B.~Menz}
\author[5]{Z.~Nunns}
\author[6]{T.~Ohnishi}
\author[5]{Zs.~Podolyak}
\author[1]{J.~Stadlmann}
\author[1, 8]{Th.~Stöhlker}
\author[3]{Q.~Wang}
\author[9]{T.~Yamaguchi}
\author[6]{Y.~Yamaguchi}
\author[3]{X.~Yan}
\author[6, 10]{A.~Yano}
\author[3]{Y.~Yu} 

\affil[1]{GSI, Darmstadt, Germany}
\affil[2]{HFHF, GSI, Darmstadt, Germany}
\affil[3]{IMP-CAS, Lanzhou, China}
\affil[4]{DESY, Zeuthen, Germany}
\affil[5]{University of Surrey, Surrey, UK}
\affil[6]{RIKEN, Wako, Saitama, Japan}
\affil[7]{University of Cologne, Cologne, Germany}
\affil[8]{Helmholtz-Institut Jena, Jena, Germany}
\affil[9]{Saitama University, Saitama, Japan}
\affil[10]{University of Tsukuba, Ibaraki, Japan}

\date{}

\begin{document}

\maketitle

\begin{abstract}
The application of software defined radio in on-line interaction with the beam processes of the heavy ion storage ring is presented. It is discussed how this new technique can enhance the beam time efficiency and open up new measurement possibilities. Discussed is a specific example to halt the accelerator running in case a rare stored particle is identified online.
\end{abstract}

\section{Introduction}
Non-destructive detection methods are often used for in-flight measurements of beam properties based on frequency analysis. In addition to typical \yl{applications in accelerator physics to acquire} beam parameters such as intensity, revolution frequency, etc., non-destructive Schottky detectors are \yl{routinely utilised as major detection systems in precision physics experiments \cite{steck2020}. Bright examples are mass and lifetime measurements of exotic nuclei in ion storage rings \cite{litvinov2011, litvinov2013, bosch2013}.}
Measurements of lifetimes as long as $\approx$ 300 days \cite{leckenby2024,sidhu2024} and as short as tens of milliseconds have been possible inside the ESR storage ring \cite{kienle2013,freire2024}.

\yl{The interest today is to address exotic nuclides lying at the outskirts of the nuclidic chart. These nuclides are inevitably short-lived and, what makes their studies very difficult, they are produced with vanishingly small rates. Therefore, to maximise the scientific output the measurements are conducted in a loop where the production of fresh ions and their investigation are done as frequently as only possible. This routine is implemented into the accelerator control system which runs continuously with a predetermined repetition cycle. However, there are cases where a very rare particle of interest is long-lived, which means that to determine its lifetime, the accelerator sequence must be abruptly halted.}

\yl{A particular physics case is the search for rare long-lived isomeric states. Isomers in the neutron-rich nuclei in the $A=180$ region were studied previously in the ESR, where among others new isomers in $^{184}$Hf and $^{186}$Hf were discovered \cite{reed2010,reed2010a}. To measure the corresponding half-lives, research team manually observed spectra and manually stopped the accelerator running. Obviously, such procedure can only be done at a repetition time of many minutes. Still man-caused errors are likely to happen. The next goal is to search for the even more exotic isomer in $^{188}$Hf, which requires a change of the procedure.}

In this work we report on the application of software defined radio in on-line interaction with the beam processes of the ESR heavy ion storage ring. We will discuss how this new technique can enhance the beam time efficiency and open up new measurement possibilities.


\section{Schottky detectors}

\yl{Resonant} Schottky cavity detectors \yl{for frequency and intensity measurements, which are often termed ``longitudinally sensitive'',} have been installed in the ESR heavy ion storage ring at GSI Darmstadt \cite{nolden2011,sanjari2013,sanjari2020}, in the CSRe storage ring at IMP Lanzhou \cite{wu2013,wang2025} as well as inside the Rare RI-Ring (R3) storage ring at RIKEN \cite{suzaki2015,yamaguchi2016}. In the latter storage ring, recently efforts have been made in order to design position sensitive \yl{(``transversely sensitive'')} Schottky cavity doublet detectors \cite{sanjari2015,chen2015,chen2016,dmytriiev2020,hudsonchang2025,sanjari2025}. 

Schottky signals consist of broadband high frequency components \cite{caspers2008}. These place high demands on the data acquisition hardware in terms of acquisition bandwidth and dynamic range. Commercially available spectrum analysers have proved very useful for various stages of experimental preparation and signal acquisition, but when it came to long-term recording of the signal, a continuous time acquisition system was required. For the ESR, originally the TCAP (time capture) system was designed \cite{kaza2004} which was followed by \yl{an upgraded} new time capture system NTCAP \cite{trageser2015}.

\section{Software Defined Radio}

One of the major drawbacks of the previous time recording systems at GSI is the high initial cost of individual modules, vendor lock and closed source operating software, which in turn is subject to a lack of long-term commercial support. This affects the reliability of these data acquisition systems and hinders their use in distributed scenarios due to lack of scalability.

\yl{On contrary,} Software Defined Radio (SDR) represents a paradigm shift in data acquisition of radio frequency signals by using software to process and manipulate signals, a task that has traditionally been handled by or hard-coded into specific hardware components. SDRs facilitate advanced signal processing techniques and enable rapid prototyping and experimentation in communications systems. They are available on the market in a variety of sizes and price \yl{span}, making them suitable candidates for both redundant and distributed data acquisition nodes.

Common to most popular SDR systems is the open source driver stack and to some extent open access hardware, making the acquisition and analysis code easily maintainable. The acquisition software itself can be published according to open science standards and F.A.I.R. (Findable, Accessible, Interoperable and Reproducible \cite{gofair}) principles, together with the other components of the research, such as raw, intermediate and result data, as well as the data analysis software.

The relevance of SDRs has already been recognised for the control system of GSI and the future FAIR project, for monitoring and acquisition of slow and fast analog signals as well as signals from different RF sources. Current versions of the GNU-Radio are now being partly developed at GSI \cite{gnuradio}.

SDRs can also be used as alternative time capture systems for Schottky detector systems for physics research in storage rings. Some progress has been made on this front in the past (see for example \cite{dmytriiev2020} or the sdrTcap project \cite{sdrtTcap}). However, the versatility of SDRs allows their use in even more \yl{sophisticated} applications in on-line monitoring and processing of acquired data.

\begin{figure}
\includegraphics[width=0.5\textwidth]{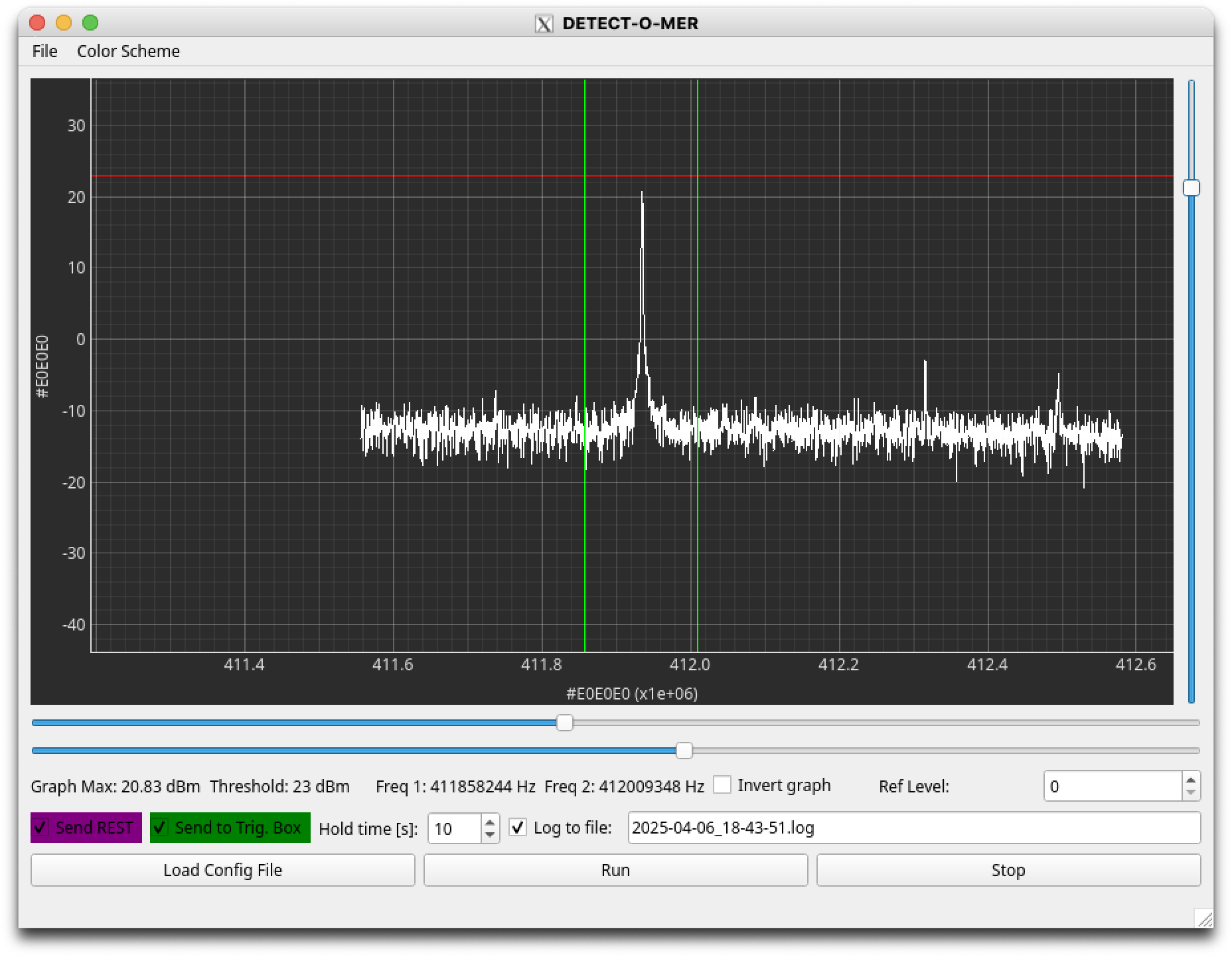}
\caption{
  The GUI interface of the code ``detectomer'' \cite{sanjari2025} for monitoring the appearance of the signal of interest: here testing the trigger on the 214th harmonic of some randomly stored fragments at 364.9 MeV/u in the ESR.}
\label{fig:gui}
\end{figure}

\section{Conditional \yl{interruption} of beam processes}

GSI's new control system allows interaction with ESR parameters and beam processes via a well-defined REST interface. For this purpose, specific blocks (also known as virtual accelerators) with unique identifiers are \yl{introduced} within beam patterns. Among the many features associated with these blocks is the ability to set a breakpoint. This would then hold the pattern, pausing the entire injection cycle. This breakpoint can be set and released manually by pressing a button on the control system GUI or by sending a message to the control system via the REST interface \cite{schaller2022}.

A new code was developed to take advantage of this feature, to make a conditional breakpoint based on the spectral power of the RF signal from the Schottky detectors \cite{sanjari2025}. This code consists of two parts in order to allow modularity and scalability for future applications.

The data sampler part has a command line interface (CLI). It acquires samples from a Realtek (RTL) SDR device according to settings in a configuration file and makes the data available on the local network as a publisher using the ZeroMQ messaging library. In this work, a \textit{NESDR SMArt v5} with larger temperature stability was used, although \yl{alternative} RTL-SDR devices can also be used.

The graphical user interface (GUI) part (see figure \ref{fig:gui}) subscribes to the data of the ZeroMQ channel on the local network and displays its Fourier spectra in real time. Zoom functions and sliders can be used to select a region where a desired peak may appear, in which case a REST signal is generated and sent in order to set a breakpoint on the beam pattern in the same beam cycle. The user can set the release time via the GUI, after which the breakpoint is reset and the beam process restarts the cycles. The GUI also allows independent logging of the exact timestamps of each event.

Previously, lifetime measurements were performed in the standard mode of operation within the ESR at fixed time intervals. For example, in the case of a two-body decay of $^{142}$Pm \yl{the aim was to store only a few ions at a time. Here,} the beam pattern was set to a fixed delay of about 64 seconds before a new injection was prepared. This time was calculated to be slightly longer than the estimated lifetime of about 42 seconds. Each injection, \yl{including inevitable the empty ones} was saved to a separate file using the spectrum analysers while the NTCAP system was running continuously in the background \cite{kienle2013}. Shorter lifetimes were measured in the isochronous mode of the ESR with much shorter cycles in the order of 10 seconds and below, looking for lifetimes in the order of milliseconds. Again, single shot acquisition was complemented by the NTCAP system running in the background \cite{freire2024}.

Common to the above measurements are the relatively short lifetimes of the species of interest on the one hand, and the high yield, i.e. production cross section, on the other. However, for species of interest with very long estimated lifetimes and very low yields, fixed duration patterns will greatly reduce the efficiency of the beam time.

\begin{figure}
  \includegraphics[width=0.5\textwidth]{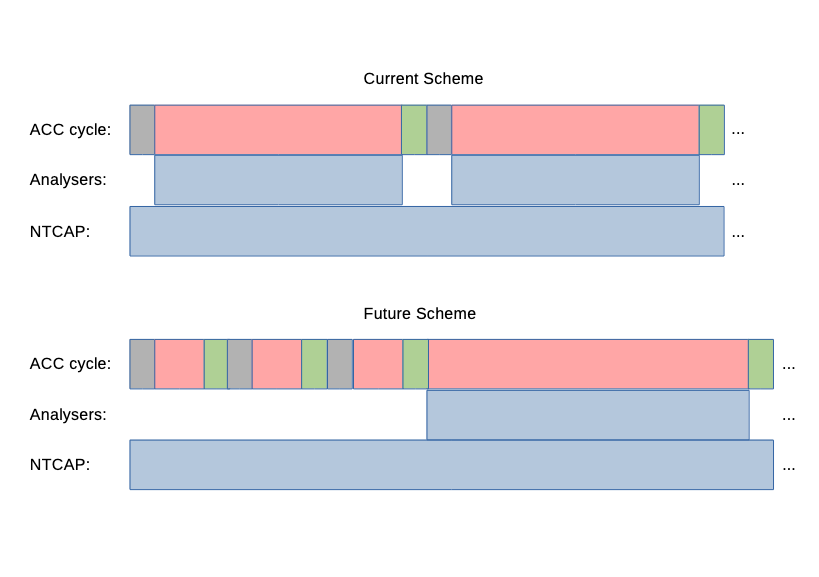}
  \caption{Schematic of the previous timing techniques in comparison to the technique suggested by this work. Gray and green boxes show accelerator cycle start (injection) and end, red boxes show accelerator waiting time and blue boxes show recording time using the triggered spectrum analysers as well as the continuously running NTCAP.}
  \label{fig:timing}
  \end{figure}

Figure \ref{fig:timing} shows schematically how the new system can greatly improve beam time efficiency by continuously monitoring the beam inside the ESR. If the \yl{rare} species of interest (e.g. a long-lived isomer) appears during the injection, it is detected and a breakpoint signal is sent to the accelerator control system. This prevents a new injection and prolongs the current beam cycle by a pre-defined time, that is set in the GUI. Otherwise a new injection is prepared within a couple of seconds.

In the previous scheme, the spectrum analysers would produce recording files of equal size. It was up to the analysis procedure to detect whether files contained an event. Also the continuous stream of NTCAP data needed to be chopped into individual segments or files using offline algorithms in order to detect the injection times. A separate code has been developed for this purpose \cite{freire2023}.

In the new scheme on the other hand, a breakpoint is set on the accelerator pattern as soon as an event is detected, and with each breakpoint, there is a timestamp entry that precisely documents the location of that event in the NTCAP stream. Also, the spectrum analysers will be triggered on that condition, making sure that every recorded injection contains an event. The triggering of the spectrum analysers is carried out using a network based distributed trigger system that was developed for these purposes \cite{rdts}. The network latency is very short but in any event does not play a major role due to the trigger offset feature of the spectrum analysers, allowing them to continuously monitor the signal using a moving time window.

\section{Summary and outlook}

Schottky \yl{cavity-based} detectors are fast and sensitive devices that can be used for mass and lifetime measurements of exotic isotopes and their isomers. Open source data acquisition systems based on Software Defined Radio complement the operation of Schottky detectors and contribute to the scalability of data handling.

Using Software Defined Radio, a new code permits, for the first time, online interaction with the accelerator beam processes and decision making during the same cycle. The suggested configuration makes it possible to address the measurement of unstable nuclide and isomer states, which are \yl{long-lived than the cycling time} but have a very low yield, thus increasing the efficiency of the beam time.

While the above code does not use the GNU-Radio library, it is envisaged that this library can be used to build even more complex \yl{and versatile} conditional systems using the combination of online beam monitoring using SDRs and coupling to accelerator parameters. The use of customised GNU-Radio library blocks would also enable connection to other GSI and FAIR infrastructure. A recent implementation now allows usage of FAIR timing system as a GNU-Radio block \cite{fairdigitizers}, this allows for precise data acquisition and synchronisation using unique time stamps.

Finally, since RTL-SDRs have a limited sampling rate, we hope to move towards SDRs with higher sampling rates and multiple inputs, such as those from Lime Microsystems \cite{lime}. Again, the GNU-Radio library would play a very important role here.

\section{Acknowledgements}

M.~S.~S., Yu.~A.~L. and J.~G. acknowledge support by the State of Hesse within the Research Cluster ELEMENTS (Project ID 500/10.006). E.~M. and Yu.~A.~L. acknowledge support by the project ``NRW-FAIR", a part of the programme ``Netzwerke 2021", an initiative of the Ministry of Culture and Science of the State of North Rhine-Westphalia.


\end{document}